\documentclass[pdflatex,sn-nature]{sn-jnl}

\usepackage{graphicx}%
\usepackage{multirow}%
\usepackage{amsmath,amssymb,amsfonts}%
\usepackage{amsthm}%
\usepackage{mathrsfs}%
\usepackage[title]{appendix}%
\usepackage{xcolor}%
\usepackage{textcomp}%
\usepackage{manyfoot}%
\usepackage{booktabs}%
\usepackage{algorithm}%
\usepackage{algorithmicx}%
\usepackage{algpseudocode}%
\usepackage{listings}%
\theoremstyle{thmstyleone}%
\theoremstyle{thmstyletwo}%

\theoremstyle{thmstylethree}%

\begin{document}

\title[Article Title]{e-Traceroute: Physically traceable electricity routing for carbon-free energy utilization}

\author[1]{\fnm{Shiu} \sur{Mochiyama}}\nomail

\author[2]{\fnm{Ryo} \sur{Takahashi}}\nomail

\author[1]{\fnm{Yoshihiko} \sur{Susuki}}\nomail

\affil[1]{\orgdiv{Graduate School of Engineering}, \orgname{Kyoto University}, \orgaddress{\street{Kyoto-daigaku Katsura}, \city{Nishikyo}, \postcode{6158510}, \state{Kyoto}, \country{Japan}}}

\affil[2]{\orgdiv{Faculty of Engineering}, \orgname{Kyoto University of Advanced Science}, \orgaddress{\street{Gotanda-cho, Yamanouchi}, \city{Ukyo}, \postcode{615-8577}, \state{Kyoto}, \country{Japan}}}

%%==================================%%
%% Sample for unstructured abstract %%
%%==================================%%

\abstract{%
Maximizing the self-consumption of residential photovoltaic generation is a promising pathway to meet urgent decarbonization targets for 2030 (and even for 2035). A viable solution is to create a sharing economy for idle battery capacity within a community. To achieve this, utilizing shared physical assets necessitates complete physical traceability of power flows---the capability to strictly trace the ownership of stored energy among multiple participants. Furthermore, physical traceability is an essential function for demonstrating the use of carbon-free energy resources. 
Such tracing is impossible in conventional systems owing to two fundamental limitations: the mixing of power flows in a common bus and the decoupling of power and information delivery. 
This study presents a novel physical-layer technology, called \textit{e-Traceroute}, that overcomes these limitations and realizes physically traceable electricity exchanges. Specifically, physically distinguishable \textit{power routing} and data transmission are unified over the same power lines. Prototyping experiments demonstrate successful integration of power transfers and information transactions, validating that the proposed system serves as a physical foundation for a reliable and scalable sharing economy that drives bottom-up decarbonization.
}


\maketitle

\section{Introduction} \label{sec:intro}

Ten years after the adoption of the Paris Agreement, global climate action is far from achieving its goals. Many countries, including G20 members, are not currently on track to achieve their Nationally Determined Contribution (NDC) targets for 2030 (and even for 2035)\cite{Programme-2025}. More fundamentally, it is argued that the current NDC targets are inadequate to limit global warming to the target temperature increase of 1.5$^\circ$C\cite{Meinshausen.etal-2022,Programme-2025,Calvin.etal-2023}. Immediate and adequate actions are required from both social and technological perspectives.

With only four years remaining until 2030, the immediate deployment of decarbonization technologies that leverage existing grid infrastructure is imperative. Building large-scale infrastructure, such as wind farms and nuclear power plants, involves long lead times\cite{Calvin.etal-2023,InternationalEnergyAgency-2019} that prevent it from meeting the near-term target. Instead, immediate solutions should rely on existing distributed renewable energy assets on a small geophysical scale (consumer side). 

An effective strategy is to increase the self-consumption of residential photovoltaic resources. Despite their rapid spread, unlocking the true potential of photovoltaic energy remains fundamentally hindered by the lack of effective means to manage surplus energy arising from mismatches of supply and load profiles under light-load conditions\cite{Masson.etal-2025,InternationalEnergyAgency-2024}. Currently, reverse power flows to the grid are highly restricted owing to the limited capacity of existing infrastructure\cite{Liang-2017,TheReliableAffordableCleanEnergyfor2030CooperativeResearchCentre-2021}. The increasing installation of residential batteries, called the \emph{behind-the-meter} batteries\cite{Rezaeimozafar.etal-2022}, has the potential to address this issue; however, affordable residential batteries usually have insufficient capacity to compensate for the entire mismatch. 

To resolve this mismatch, establishing a \textit{sharing economy} among prosumers through peer-to-peer electricity trading\cite{Pena-Bello.etal-2022,Parag.Sovacool-2016} can be a pivotal strategy. Specifically, a sharing economy of distributed batteries between networked households\cite{Kalathil.etal-2019,Walker.Kwon-2021} introduces a platform to pool and trade the unused capacity of existing batteries, rather than requiring individual households to make isolated investments in large-capacity batteries. This collective approach is expected to maximize the overall self-consumption of photovoltaic energy across the community while encouraging participation through financial incentives.

To practically implement such a battery sharing economy within actual power distribution grids, ensuring the \textit{traceability} of power flows---strictly verifying their physical origin and destination---is of vital importance. This capability is essential for determining exactly which portion of the power belongs to whom, implying physically accurate metering of how much carbon-free energy is generated and consumed, which is completely different from virtual trading. However, this physical tracking is difficult to realize because power flows inherently mix within a shared network infrastructure. Consequently, establishing absolute physical-layer traceability poses a critical challenge that must be overcome to secure the reliability and validity of the sharing economy.

Furthermore, scaling this sharing economy to a community with numerous participating prosumers presents a primary technical challenge. Intuitively, building an energy management system appears most viable under a centralized architecture; relying on a single, omniscient coordinator with absolute visibility into the real-time state of each node can significantly simplify the information system architecture. However, as the number of participating prosumers scales, the centralized architecture suffers from a fatal bottleneck of massive data aggregation and system-wide optimization, resulting in non-negligible communication and processing delays. For a system based on shared physical infrastructure, any discrepancy between computational decisions and actual physical power flows is fundamentally unacceptable; such control lags invalidate traceability and render electricity ownership entirely unidentifiable. Without physical traceability, any electricity exchange inevitably degrades into mere virtual trading. Crucially, virtual trading of carbon-free energy without physically accurate metering is not a definitive solution to curbing global warming.

This study develops a physical-layer technology, called \textit{e-Traceroute}, that establishes a hardware-level coupling between power and information flows, effectively overcoming the fundamental limitations in the traceability of electricity.
Analogous to the network diagnostic tool that traces data packets across IP networks, e-Traceroute guarantees the strict, \textit{physical} tracking of  electricity.
Specifically, a decentralized, edge-oriented architecture is constructed that closely coordinates \textit{power routing} with its associated information processing and communication. This integrated routing paradigm is introduced to dynamically control physical power flows by forging a verifiable one-to-one link between the intended and actual power flow.
To physically realize this integration, power routing is co-designed based on circuit switching\cite{Takuno.etal-2010,Takuno.etal-2011b} and power line communication (PLC)\cite{Galli.etal-2011} into a unified medium for both power transfers and information transactions. 
Among the emerging technologies for power and information integration\cite{He.etal-2020,Hoeher.etal-2025,He.etal-2024,Zhang.Ho-2013}, combining power routing and PLC offers the practical route to leveraging existing battery infrastructure without modifications. 
Experiments using a multi-port prototype hardware demonstrate that power and information can be routed concurrently as an inseparable entity, thereby strictly guaranteeing information integrity and the subsequent traceability of power flows. By ensuring that the intent of a power transfer and its physical execution are inherently inseparable, these findings establish a reliable and scalable foundation for a fully accountable sharing economy.

\section{Results}

\subsection{Integrated architecture for traceable power routing}

\begin{figure}[tb]
    \centering
    \includegraphics[width=\linewidth]{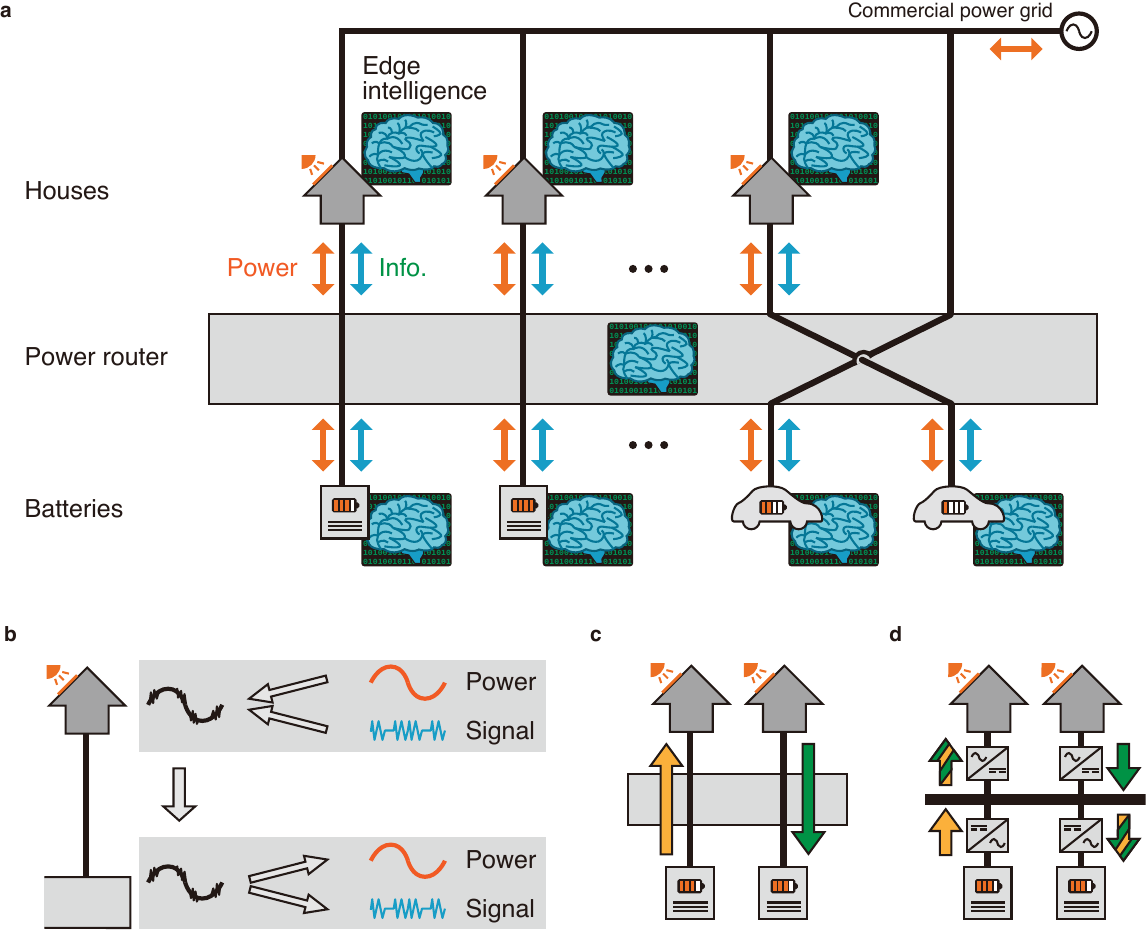}
    \caption{\textbf{Energy management system (EMS) based on integrated power transfers and information transactions. }
    \textbf{a}, Overview of the EMS. The lines inside the power router represent a specific example of the one-to-one relationships between prosumers. 
    \textbf{b}, Integration of power transfers and information transactions through power line communication technology. 
    \textbf{c~and~d}, Comparison of the proposed routing-based power transfers (c) and converter-based counterpart (d). The former ensures the separated power flows between two transfer pairs while the latter, the conventional converter-based system inevitably mixes power flows within a common bus. 
    }
    \label{fig:overview}
\end{figure}

Fig.~\ref{fig:overview}a illustrates the target energy management system (EMS). The EMS manages power transfers between prosumers, including houses with photovoltaic generation and stationary or vehicle-mounted batteries. The key enabler of traceable power transfers over the proposed system is the integration of power routing and PLC (Fig.~\ref{fig:overview}b). The router hardware forms a switch matrix that dynamically establishes a one-to-one connection between two specific prosumers \cite{Takuno.etal-2011b,Takahashi.etal-2013a}. Parallelizing these one-to-one connections enables concurrent, distinguishable power transfers across the entire system (Figs.~\ref{fig:overview}c~and~d). Unifying these power transfers with their associated information transactions via PLC guarantees strict traceability of exactly which portion of the stored energy of the batteries belongs to whom.

For the control of the EMS, an edge-oriented architecture with distributed intelligence \cite{Cheng.etal-2018,Choi-2019} is employed. Each edge prosumer autonomously executes local optimization, such as forecasting domestic net-load profiles and determining individual battery scheduling, using its own real-time data to maximize local utility. Simultaneously, the central router performs system-wide coordination and determines the state of the switching matrix based on dimensionally reduced data aggregated from the edges. This functional decentralization effectively relieves the central router from a prohibitive computational burden and eliminates the communication capacity bottleneck, thereby securing the scalability of the proposed EMS. 

The physical integration of power routing and information transactions plays an essential role in practically operating this decentralized architecture. The traceability of power flows is contingent on exactly synchronized operation of spatially distributed edge prosumers and the centrally located power router. The target actions derived from the optimization must be shared between the edge prosumers and power router. The circuit switching of the router and the control of the edge batteries must coincide. Any temporal mismatch between them results in unverified power flows, invalidating the very premise of physical traceability. Tight synchronization of power and information is achieved by leveraging the direct integration of PLC into the physical hardware of both the power router and the prosumer nodes. This embedded communication enables the EMS to execute the physical circuit switching at the central router and the charge/discharge operations at the edge batteries without temporal mismatch. 

The integrated PLC handles the two essential types of information exchange within the distributed architecture: the asynchronous aggregation of locally processed edge data and the timing-critical multicast of control commands. Although data aggregation has relaxed timing requirements and could technically rely on commercial cellular networks, PLC is adopted to guarantee robust and autonomous operation even during unpredictable network failures or complete outages, such as those caused by natural disasters. Furthermore, and more fundamentally, the uncontrollable latencies of an external medium are completely unacceptable for the control commands. Such delays inevitably disrupt the synchronization between the router and edges. Unifying the entire communication system over the PLC makes the communication path physically identical to the power transmission path, entirely bypassing external network delays. Consequently, information acts as a verifiable physical proxy for the associated electricity. 

The integration of power routing and PLC offers a highly practical pathway to address the urgent decarbonization target. While the emerging talkative power approach \cite{He.etal-2020,Hoeher.etal-2025} offers a hardware-saving alternative for power-and-information integration by coordinating data transmission into the control loops of power electronic converters, this approach requires internal modifications to the equipment. This presents a practical hurdle for the specific application, which strictly assumes the use of existing residential batteries without altering their proprietary pulse width modulation generation or safety certifications. Although simultaneous wireless information and power transfer \cite{He.etal-2024,Zhang.Ho-2013} is a promising approach in ensuring coincidence of energy and data, scaling wireless power distribution to residential power ratings and maintaining a high conversion efficiency remain challenging. Ultimately, the proposed method is best suited for developing an EMS that leverages existing edge assets without internal modifications.

\subsection{Experimental validation of integrated power transfers and information transactions}
\begin{figure}[!t]
    \centering
    \includegraphics[width=.6\linewidth]{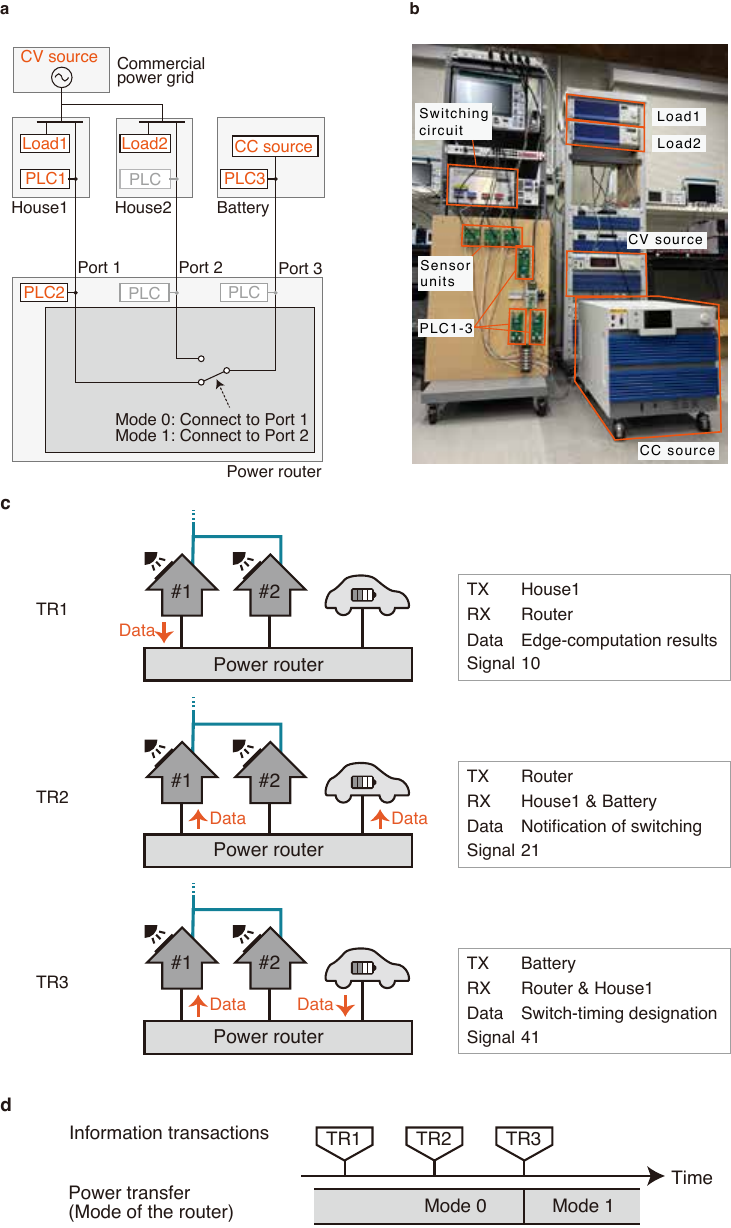}
    \caption{\textbf{Experimental validation setup.} \textbf{a}, Overall configuration. The power line communication (PLC) modules greyed out in the figure are omitted in the setup because they are not necessary in the scenario of interest. 
    \textbf{b}, Photograph of the setup. 
    \textbf{c}, Information transaction scenario. 
    \textbf{d}, Timeline of power routing and information transactions in the scenario. }
    \label{fig:setup}
\end{figure}

A hardware prototype was developed consisting of a power router and three edge prosumers, each equipped with the integrated PLC modules. Figs.~\ref{fig:setup}a~and~b present the setup for experimental validation (see Methods for hardware details). 
The setup captures a subsystem of the entire EMS, focusing on two houses and one battery. 

A dynamic power-routing scenario was designed that sequentially executes the two essential types of information exchange through a three-step communication process. Figs.~\ref{fig:setup}c~and~d present the experimental verification scenario. First, for asynchronous data aggregation, an edge house sends locally processed data to the router to request a new power transfer (TR1). Subsequently, the router determines the next connection and multicasts a switching notification to the edge prosumers (TR2). Finally, the battery multicasts a timing-critical command of circuit switching (TR3). The operating mode of the router is determined based on its receiving signals. In this scenario, the transmission of signal 41 from the battery to the router in TR3 triggers the transition from Modes 0 to 1. 

To highlight the timing-critical nature of the last information transaction and its subsequent power transfer, a zero-power switching strategy was implemented. By design, the arrival of this command at the router immediately triggers physical circuit switching. The transmission of this command was then set to occur exactly at the zero-crossing point of the instantaneous power. The zero-power switching provides a practical advantage of mitigating the adverse effects of switching transients, such as overvoltage surges on devices and electromagnetic interference, as well as visualizing the synchronized power transfers and information transactions. 

\begin{figure}[!t]
    \centering
    \includegraphics[width=0.85\linewidth]{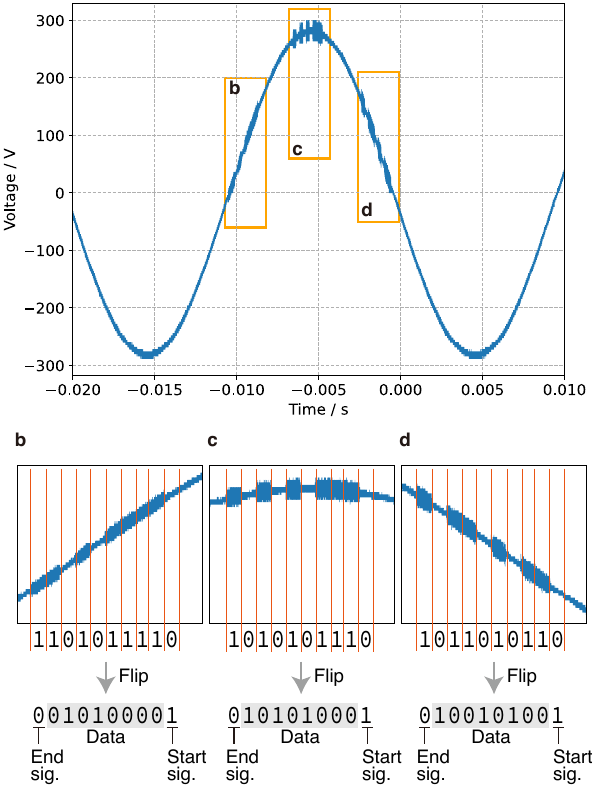}
    \caption{\textbf{Voltage waveforms of the integrated power transfers and information transactions.}
    \textbf{a}, Full-frame view of the voltage waveform around the information transactions. 
    \textbf{b, c, and d}, Enlarged views of the voltage waveform around each information transaction. Each part corresponds to the rectangles indicated by b, c, and d in a. The intervals of the thin vertical lines in orange represent the 1-bit duration of the signal. The numbers below the waveforms indicate the corresponding logic, 0 and 1. By design, the data payload consists of a single-byte signal between a start signal and an end signal and is logically inverted (See Methods for the detailed protocol). Note that the time $0\,\mathrm{s}$ does not coincide between this figure and the others (Fig.~\ref{fig:result_signals}--\ref{fig:result_power}). }
    \label{fig:result_volt}
\end{figure}

Following this scenario, the transmission (TX) operation of the developed modules is first confirmed.
Fig.~\ref{fig:result_volt} presents the voltage waveform measured at Port 3 of the router, capturing the three sequential information transactions. 
Figs.~\ref{fig:result_volt}~(b)--(d) present an enlarged view of the three transactions. 
In Fig.~\ref{fig:result_volt}~(b), the waveform indicates that the signal is ``0010100001." 
The first bit is the start signal, and the following eight bits represent the data in the least-significant-bit-first manner. 
Thus, the data are represented as binary values, e.g., ``00001010" denotes the decimal number 10. 
In the same way, the signals in Fig.~\ref{fig:result_volt}~(c)~and~(d) are read as ``00010101" (21 in decimal) and "00101001" (41 in decimal), respectively. 
These values successfully encode the respective signals for the three-step scenario: asynchronous data aggregation, switching notification, and timing-critical trigger command. Therefore, the PLC modules correctly imposed the intended information sequences onto the power waveforms.

\begin{figure}[!t]
    \centering
    \includegraphics[width=0.95\linewidth]{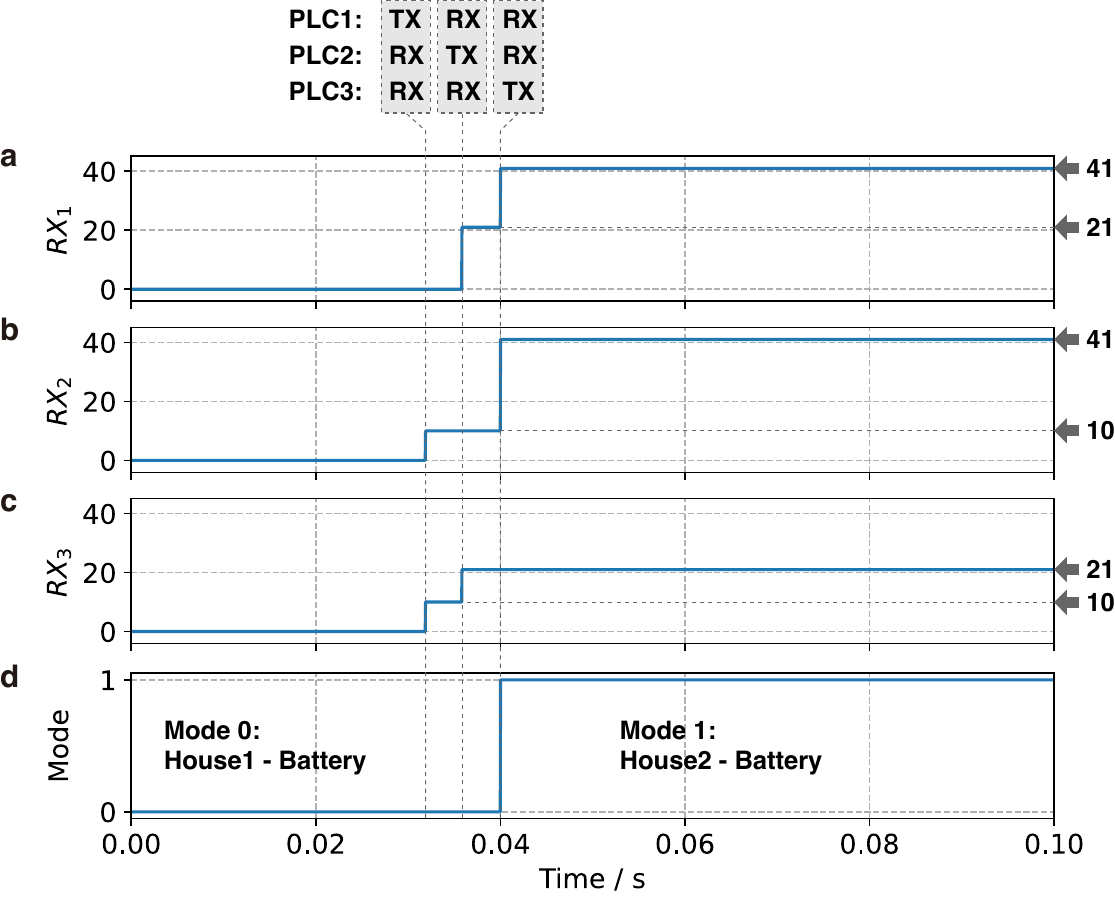}
    \caption{\textbf{Transitions of the RX register values (a, b, and c) and the mode of the router (d)}. RX$_{1}$, RX$_{2}$, and RX$_{3}$ denote the register values of PLC modules 1, 2, and 3, respectively. }
    \label{fig:result_signals}
\end{figure}

The receiver (RX) operations and the subsequent hardware responses are then confirmed. Fig.~\ref{fig:result_signals} shows the transitions of the received signals in the three modules along with the operating mode of the router. 
The mode represents the internal variable of the router whose value corresponds to one of the switching states. As explained previously, the mode transition was caused by $RX_2$ changing to 41. 
The RX signals synchronously transitioned through 10, 21, and finally 41, which demonstrates that the PLC modules successfully delivered the information across the power line. Note that half-duplex communication was adopted; therefore, the module in the TX mode did not renew its register value in the corresponding transactions. Most importantly, the mode of the router then changed from 0 to 1 at the exact moment when the register value of the PLC module of the router transitioned to 41. This result verifies that the router instantly executed the physical operation driven by the received timing-critical signal.

\begin{figure}[!t]
    \centering
    \includegraphics[width=0.95\linewidth]{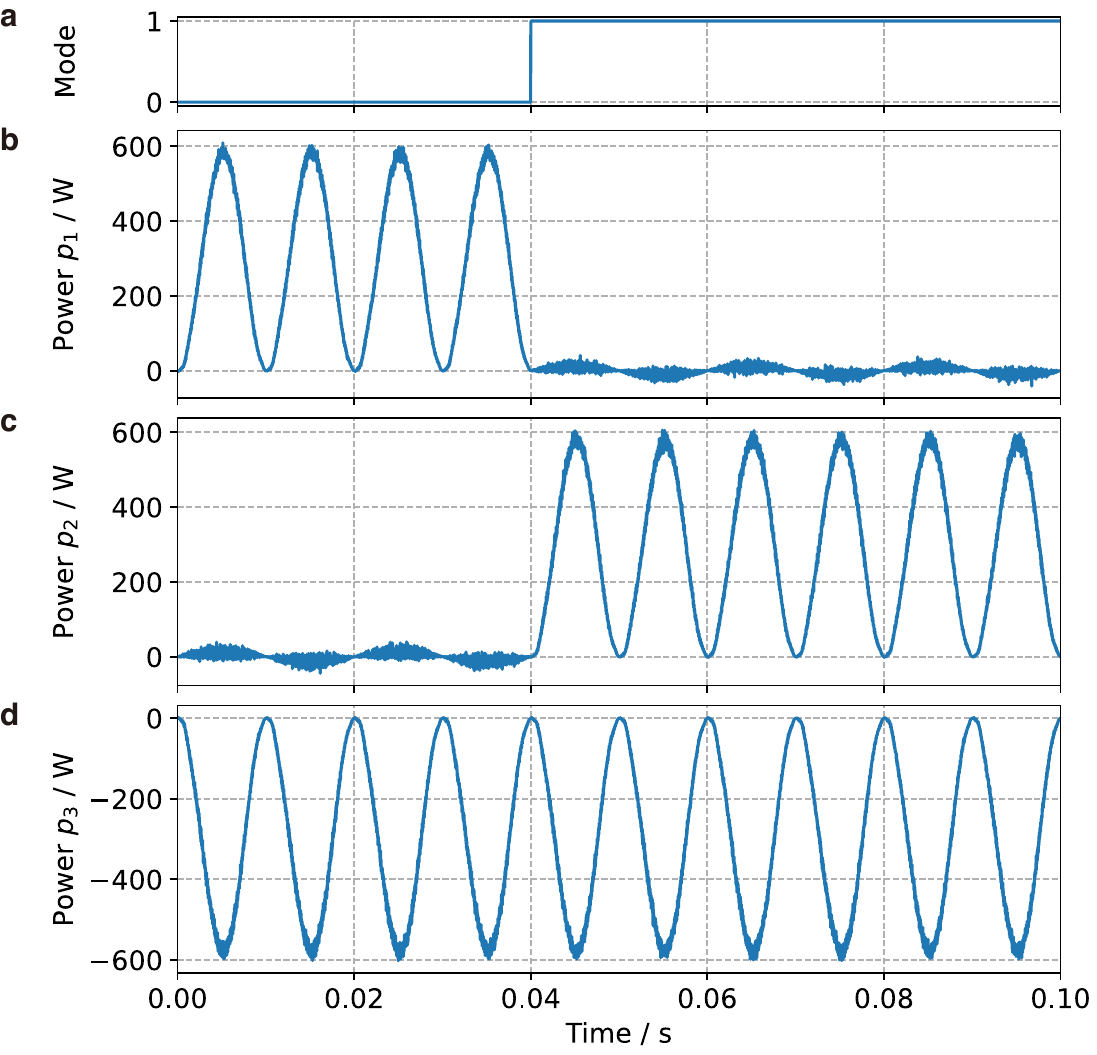}
    \caption{\textbf{Power transfers around the mode change of the router. } 
    \textbf{a}, Transition of the mode of the router (reproduced from Fig.~\ref{fig:result_signals}d). 
    \textbf{b, c, and d}, Instantaneous power measured at Ports 1, 2, and 3, respectively. Power flowing out of the router is defined as positive. }
    \label{fig:result_power}
\end{figure}

Lastly, the resulting physical power routing was evaluated. Fig.~\ref{fig:result_power} presents the power waveforms measured at the ports of the router along with the transition of the mode of the router. 
The waveforms $p_1$, $p_2$, and $p_3$ represent the instantaneous power measured at Ports 1, 2, and 3, respectively. 
The destination of the power supply from the battery transitioned seamlessly from Houses 1 to 2 immediately after the router mode changed. Most importantly, this physical switching of the routing pairs occurred when the instantaneous power was at zero. This result validates that the hardware-level coupling successfully eliminates uncontrollable communication delays and thereby achieves the switching exactly at the zero crossing. From these results, it is concluded that power routing was successfully controlled by the information conveyed over the same physical lines.

\section{Discussion}

This study developed and experimentally validated a novel hardware foundation for integrating power transfers and information transactions via a line-switching-based power router. 
By utilizing PLC to convey information directly over the physical power transmission paths, hardware-level coupling was achieved between the timing-critical signal and power flow. 
Experimental results using a multi-port prototype hardware demonstrated the coordinated control of the power routing and the information transactions. 

This physical integration addresses the fundamental limitation of conventional information-based management systems, where the decoupling of data and power flows hinders absolute traceability. 
The proposed method ensures that the intent of a power transfer and its actual execution are physically inseparable, providing a robust foundation for active control of power flows based on distributed architecture for information processing. 

The physical traceability relies strictly on the separation of power flows (Fig.~\ref{fig:overview}b). This fundamental requirement clearly distinguishes the proposal from existing concepts related to power routing. Conventional power routing is primarily based on power electronic converters (e.g., \cite{Huang.etal-2011,Abe.etal-2011,Stalling.etal-2012,Kado.etal-2016,Yu.etal-2023,Liu.etal-2023,Nie.etal-2025}). Inherent to the nature of power conversion, such routers mix input power from different origins on a common DC or AC bus and distribute it by proportional division (Fig.~\ref{fig:overview}c). To overcome this intrinsic limitation in establishing an accountable sharing economy, power transfers must be executed via physically isolated closed circuits.

The concept of physical power routing originated in the 2010s \cite{Takuno.etal-2010}. Since its inception, is has evolved into two distinct trajectories: power packetization targeting DC power systems \cite{Takahashi.etal-2015,mochiyamaPowerPacketDispatching2021,Mochiyama.etal-2025}, and AC power routing designed for distribution networks. As the primary objective is to leverage existing grid infrastructure without invasive modifications, this study specifically focused on AC power routing. Historically, line-switching power routers were first utilized for intra-house power management \cite{Takuno.etal-2011b,Takahashi.etal-2013a}. More recently, we extended the capabilities of these routers in terms of power ratings and bi-directionality toward community-level energy management \cite{Mochiyama.etal-2024a,Mochiyama.etal-2025a}. 

However, realizing true physical traceability in a distributed system requires more than just circuits; it necessitates flawless synchronization between the physical power switching and associated information processing. This study makes a pivotal contribution by providing the missing link: introducing a hardware-level integration of power and information. While earlier studies\cite{Takuno.etal-2011b,Takahashi.etal-2013a} employed PLC to send unidirectional and simple experimental commands to the router, their focus remained solely on the switching operations themselves. Architectural details addressing the tight coupling of power and communication over distributed nodes were completely unexplored. This study addresses this gap by proposing a fully distributed architecture, implementing its practical hardware, and experimentally demonstrating the concurrent power and information transmission. Most notably, eliminating uncontrollable network latencies through this unified physical medium establishes a technical foundation that is far beyond the scope of any existing technologies. 

Future research will focus on the development of higher-layer protocols for optimizing multi-agent electricity exchanges built upon the physical foundation of traceable power routing. This will unlock the potential of the reliable and accountable sharing economy to significantly advance the decarbonization. 

\section{Methods}

\subsection{Hardware implementation of the power router}

\begin{figure}
    \centering
    \includegraphics[width=\linewidth]{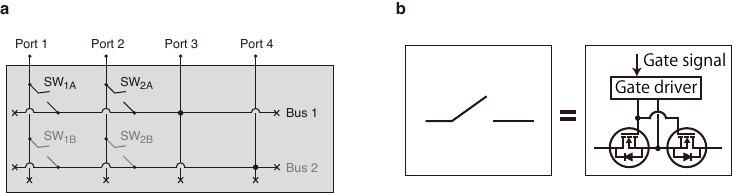}
    \caption{\textbf{Hardware configuration of the power router.} 
    \textbf{a}, Switching circuit consists of a switching matrix with four ports and two buses. 
    \textbf{b}, Each switch is a back-to-back connection of two MOSFETs. } 
    \label{fig:hardware}
\end{figure}

Fig.~\ref{fig:hardware} shows the hardware configuration for the prototype power router developed in our prior work\cite{Mochiyama.etal-2024a,Mochiyama.etal-2025a}. We used the same configuration in the experiments in this study. 
The router circuit comprises a crossbar-type switching matrix. 
The vertical lines represent the ports, which connect to edge prosumers. 
The horizontal lines correspond to the internal buses. 
The number of buses corresponds to the maximum number of simultaneous routing pairs. 
The bidirectional switches are back-to-back connections of two SiC metal-oxide-semiconductor field-effect transistors (MOSFETs) of $650\,\mathrm{V}$, $20\,\mathrm{A}$ ratings (C3M0060065; Wolfspeed, USA). 
Turning on a switch connects the edge prosumers to a horizontal line, though which the prosumers form a routing pair. 
This matrix structure ensures that routing pairs are formed over physically isolated horizontal buses, thereby preventing the mixing of power flows from different sources.

In the experiment, we used only switches SW$_\mathrm{1A}$ and SW$_\mathrm{2A}$ (presented in thicker colour in Fig.~\ref{fig:hardware}a) because power transfers between Ports 1--3 were considered and the maximum number of routing pairs was one. 
Modes 0 and 1 of the router, defined in the Results section (Fig.~\ref{fig:setup}a), correspond to the switching states $(\mathrm{SW}_\mathrm{1A},\mathrm{SW}_\mathrm{2A})=(\mathrm{ON},\mathrm{OFF})$ and $(\mathrm{SW}_\mathrm{1A},\mathrm{SW}_\mathrm{2A})=(\mathrm{OFF},\mathrm{ON})$, respectively. 

\subsection{Hardware implementation of the PLC module}

\begin{figure}
    \centering
    \includegraphics[width=0.9\linewidth]{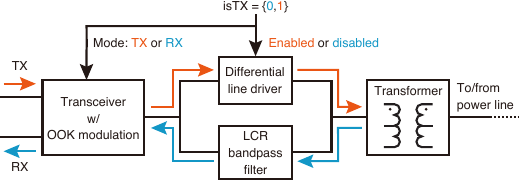}
    \caption{
    \textbf{Functional block diagram of the PLC module. }
    Parts in orange and blue correspond to the functions as a transmitter (TX) and receiver (RX), respectively. The functions are switched by the external signal (isTX). }
    \label{fig:plcmodule}
\end{figure}

PLC modules serve as converters between a digital signal handled by the controllers and a modulated voltage waveform on the power line. 
PLC modules were fabricated based on a reference design\cite{TexasInstruments-2025} with some adaptations to suit the required ratings. 
Fig.~\ref{fig:plcmodule} illustrates the functional block diagram of the PLC module. 
The module operates in one of two modes, TX and RX, because half-duplex communication is implemented. 
The mode is determined by external signals.
In the TX mode, the transceiver integrated circuit (IC) passes the digital signal modulated in on-off keying (OOK) to the differential line driver. 
In the RX mode, the transceiver IC receives the signals from the parallel-connected bandpass filter circuit and the differential line driver IC is disabled. 
The TX and RX signals are produced or read by a controller (see Section~\ref{ssec:software} for a detailed description of the algorithm). 
The transceiver IC produces a high frequency signal while its input (TX) is 0. 
The carrier frequency of the OOK modulation was set as $125\,\mathrm{kHz}$. 
The RF transformer between the power line and the driver output and filter input signals facilitates galvanic isolation and common-mode noise rejection. 

\subsection{Software implementation of the PLC module} \label{ssec:software}

\begin{figure}
    \centering
    \includegraphics[width=.8\linewidth]{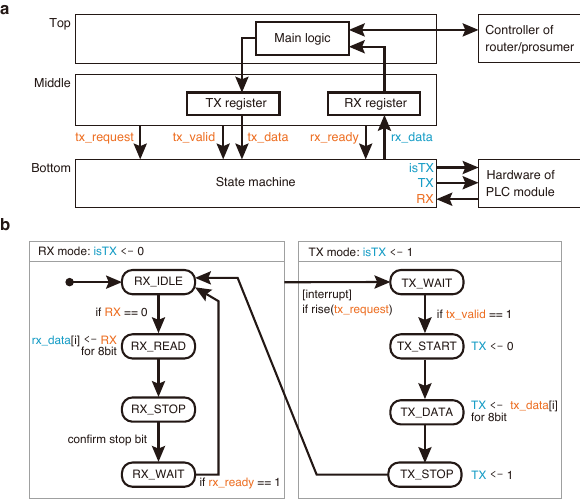}
    \caption{
    \textbf{Overview of PLC module controller. }
    \textbf{a}, Overview of the whole architecture. \textbf{b}, Simplified state transition diagram for the state machine depicted in a. }
    \label{fig:plc_controller}
\end{figure}

Fig.~\ref{fig:plc_controller}a presents an overview of the PLC module controller, which is divided into three layers, presented as top, middle, and bottom. 
The top layer determines the signal to be transferred by PLC based on the value of the received data and energy management strategies of the edge prosumers or the router connected to the module. 
The bottom layer is responsible for the actual control of the hardware according to the command from the top layer. 
A universal asynchronous receiver/transmitter-like protocol was adopted for the implementation of the bottom layer. 
The middle layer interfaces the top and bottom layers using registers and its peripheral components. 

Fig.~\ref{fig:plc_controller}b depicts a simplified state transition diagram of the state machine, which is the main component of the bottom layer. 
For half-duplex communication, the controller is by default in RX mode and enters TX mode only when an interrupt signal (tx\_request) is asserted. 
After transferring tx\_data, the state returns to RX\_IDLE and awaits an incoming signal. 
A 10-bit configuration was set for TX, comprising a 1-bit start signal, 1 byte of data, and a 1-bit stop signal. 
Specific data payloads of 10, 21, and 41 (in decimal) were used, corresponding to data aggregation, switching notification, and timing-critical trigger, respectively. 
Note that these values hold no specific significance within the protocol; they are simply arbitrarily chosen 1-byte integers. 
The Baud rate was set at 5000, namely the 1-bit length was $200\,\mu\mathrm{s}$. 

% explain the top layer
The top-layer main logic for House 1, the router, and the battery was implemented as follows to coordinate the synchronized delivery.
The experimental sequence was initiated by House 1, which transmitted a data payload of 10 and subsequently entered the RX mode to monitor incoming data. 
The router was initially idling in the RX mode. On detecting a payload of 10, it introduced a 2\,ms delay to simulate data aggregation, controller communication, and processing overhead. Following this delay, the router transmitted a data payload of 21 and reverted to the RX mode. It then instantly issued a switching command to the router controller upon receiving a timing-critical signal with a payload of 41.
Simultaneously, the battery remained in the RX mode until it detected a payload of 21, which triggered the phase-monitoring mode to track the AC voltage phase via the battery controller. This phase detection was implemented via a second-order generalized integrator phase-locked loop \cite{Ciobotaru.etal-2006}. To guarantee zero-power switching, the transmission start timing was precisely calibrated so that the end of the transmission coincided with the zero-crossing points. The zero-crossing occurred when the cosine-based phase estimation reached $\pi/2$ or $3\pi/2$. Given the 2\,ms duration of the 10-bit information transaction, the onset of transmission was advanced by $2\pi\times50\times2\times10^{-3}\,\mathrm{rad}$ relative to the target zero-crossing.

\subsection{Experimental setup of prosumers} \label{ssec:scenario}

In the experiments, the power supply and consumption were represented by regulated power sources and electronic loads. 
A constant-voltage (CV) source (PCR100LE; Kikusui Electronics, Japan) simulated a commercial power grid of $200\,\mathrm{Vrms}$, $50\,\mathrm{Hz}$. 
Two electronic loads (PCZ1000A; Kikusui Electronics, Japan) simulated the power consumptions of the corresponding houses of $500\,\mathrm{W}$. 
Power consumption was defined as total load minus PV generation. 
The regulated constant-current (CC) source (PCR6000WEA2R; Kikusui Electronics, Japan) simulated the battery with a $1.5\,\mathrm{A}$ output. 

\section*{Acknowledgements}

The authors thank Professor Takashi Hikihara of Kyoto University for his invaluable advice.

\end{document}